\documentclass{ifacconf}

\usepackage{amsmath}        
\usepackage{amssymb}        
\usepackage{diagbox}        
\usepackage{float}          
\usepackage{graphicx}       
\usepackage{verbatim}       
\usepackage{multirow}       
\usepackage{threeparttable} 
\usepackage{subfigure}         
\usepackage{enumitem}       
\usepackage{natbib}         
\newtheorem{theorem}{Theorem}
\newtheorem{lemma}{Lemma}

\newtheorem{assumption}{Assumption}
\newtheorem{problem}{Problem}

\begin{document}
\begin{frontmatter}

\title{Networked pointing system: Bearing-only target localization and pointing control} 

\thanks[footnoteinfo]{This work is supported by National Natural Science
Foundation of China (Grant No.62373386).}

\author[First]{Shiyao Li} 
\author[Third]{Bo Zhu}
\author[First]{Yining Zhou}
\author[First,Second]{Jie Ma}
\author[First,Second]{Baoqing Yang} 
\author[First,Second]{Fenghua He} 

\address[First]{
Control and Simulation Center,
Harbin Institute of Technology, Harbin, 150001, China, (e-mail: majie@hit.edu.cn)}
\address[Second]{National Key Laboratory of Complex System Control and Intelligent Agent Cooperation, Beijing, 100074, China}
\address[Third]{Advanced Control and Smart Operations, Nanjing University, Suzhou, 215163, China}

\begin{abstract}                
In the paper, we formulate the target-pointing consensus problem
where the headings of agents are required to point at a common target. Only a few agents in the network can measure the bearing information of the target. 
A two-step solution consisting of a bearing-only estimator for target localization and a control law for target pointing is constructed to address this problem. 
Compared to the strong assumptions of existing works,  we only require two agents not collinear with the target to ensure localizability. 
By introducing the concept of virtual fusion node, we prove that both the estimation error and the tracking error converge asymptotically to the origin. The video demonstration of the verification can be found at 
{https://youtu.be/S9-eyofk1DY}.

\end{abstract}

\begin{keyword}
Networked control, cooperative localization, target pointing.
\end{keyword}

\end{frontmatter}

\section{INTRODUCTION}
\begin{figure}[t]
	\centering  
\subfigure[Optical communication.]{\includegraphics[width=0.475\linewidth]{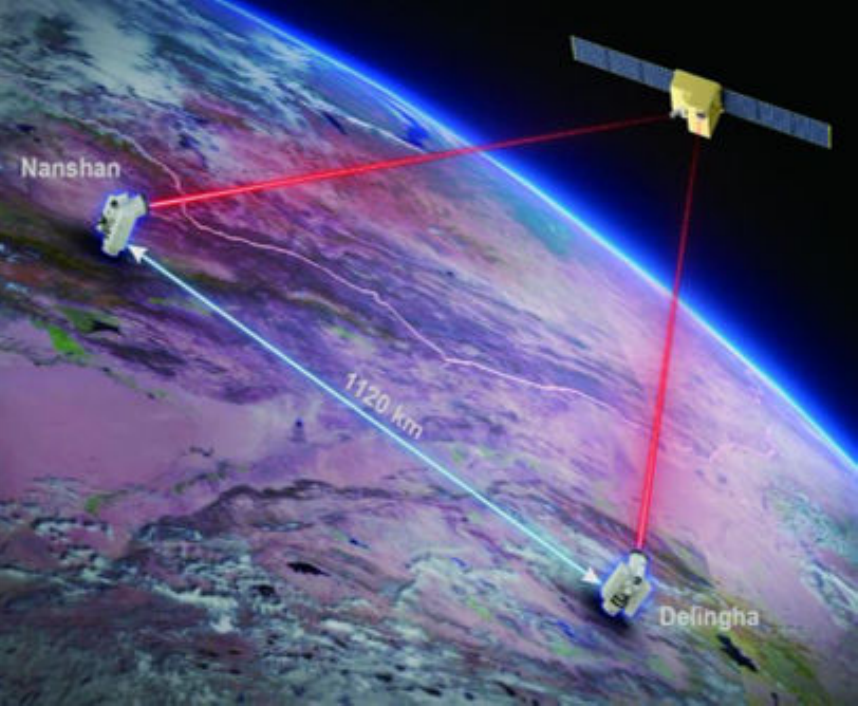}}
\subfigure[Space observation.]{
    \includegraphics[width=0.475\linewidth]{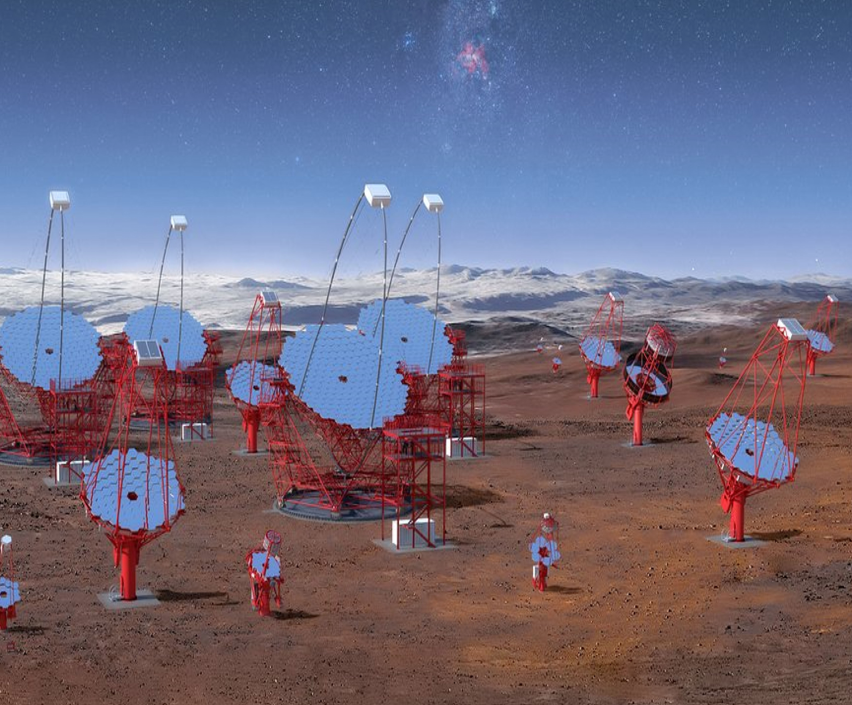}}
	\caption{The applications of networked pointing systems.}
    \label{pic_scenarios}
\end{figure}

As a growing trend, the networked pointing control technique has received remarkable attention due to its important applications, such as optical communication (\cite{yin2020entanglement}) and cooperative space observation (\cite{dubus2013surveys}). 
In the above tasks, sensors or actuators are required to point at targets for supporting distributed tracking and observation (Fig.  \ref{pic_scenarios}). These requirements promote the emergence and development of target-pointing consensus. 
In the typical studies,
to achieve the distributed pointing consensus, agents are generally required to be arranged into a prescribed  formation. 
\cite{wu2023cooperative} proposed a distributed pointing controller under a strong assumption that the agents are collinear. 
Although this assumption is removed in (\cite{trinh2018pointing}), this method requires much priori
information of target, including the desired heading vector for a leader agent, and several subtended angles for the other agents. To the best of our knowledge, constructing a distributed pointing controller without collinearity assumption remains an open issue.

Some works employ the two-step strategies, where the agents will cooperatively locate  the target  first. 
For the issue of target localization,  
solutions utilizing the Kalman filter (\cite{li2022three}) and the subspace approach (\cite{luo2019novel}) are developed. However, the computational complexity of these methods is high. Different from the previous solutions, \cite{deghat2014localization} and \cite{chen2023cooperative} constructed estimators through geometric relationships. Although the computational complexity of these estimators is lower, the localizability of these estimators relies on the persistently exciting (PE) condition. 
Motivated by the above observations, we focus on the  target-pointing consensus problem of a multi-agent system in this paper. The main contributions of this paper include the following aspects:
\begin{itemize}

\item We provide a distributed bearing-only estimator for target localization. This estimator supports all agents in the network to accurately locate targets, even if some agents cannot obtain the bearing information. 

\item A two-step strategy consisting of target localization and pointing control is proposed to achieve target-pointing consensus. This strategy gets rid of the collinearity assumption, which means that agents don’t need to be arranged into a specific formation.

\item Based on a hierarchical topology graph with three layers, we prove that both target-position estimation errors and target-bearing tracking errors can  converge asymptotically to $\boldsymbol{0}$. We also perform two simulation cases to verify the effectiveness of our strategy.

\end{itemize}

\section{PROBLEM FORMULATION}\label{section_p}
Consider a network with $n$ agents in 2-D Euclid space.  The position $\boldsymbol{p}_i \in R^2$ of each agent $i$  ($i\in\mathcal{I}\triangleq\{1,2,\ldots,n\}$) are stationary ($\boldsymbol{\dot{p}}_i=\boldsymbol{0}$), and the location is known by itself. 
Let the $\left \| \cdot  \right \|$ be the Euclidian norm of a
vector or the spectral norm of a matrix.
The heading of agent $i$ is described by a unit vector $\boldsymbol{h}_i \in R^2$, $\left \|\boldsymbol{ h}_i \right \| =1$. The relationship between heading $\boldsymbol{h}_i$ and Euler angle is
\begin{eqnarray}\label{model1}
{\boldsymbol{h}_{i}}={{\left[ \cos {{\varphi }_{i}},\sin{{\varphi }_{i}} \right]}^{T}},
\end{eqnarray}
where ${{\varphi }_{i}}\in \left[ 0,2\pi  \right)$ denotes the Euler angle of agent $i$. The end point of heading vector $\boldsymbol{h}_i$ is given by
\begin{eqnarray}\label{model2}
{\boldsymbol{p}_i}'={\boldsymbol{p}_i}+{\boldsymbol{h}_i}.
\end{eqnarray} 
The orthogonal projection matrix corresponding to $\boldsymbol{h}_i$ in 2-D Euclid space is defined as
\begin{eqnarray}\label{model3}
{\boldsymbol{M}_{{{h}_{i}}}}={\boldsymbol{I}_{2}}-{\boldsymbol{h}_{i}}{\boldsymbol{h}_{i}}^{T}.
\end{eqnarray}
Based on the orthogonal projection matrix, the dynamics of ${\boldsymbol{p}_i}'$ is given by
\begin{eqnarray}\label{model4}
{\boldsymbol{\dot{p}}_i}'=\boldsymbol{u}_i = {\boldsymbol{M}_{{{h}_{i}}}} \boldsymbol{c}_i,
\end{eqnarray}
where $\boldsymbol{c}_i \in R^2$ is the control signal that needs to design.  The change rate of ${\boldsymbol{h}_{i}}$ is derived by
\begin{eqnarray}\label{model5}
{\boldsymbol{\dot{h}}_{i}}=\frac{d}{dt}\left( \frac{{\boldsymbol{p}_{i}}'-{\boldsymbol{p}_{i}}}{\left\| {\boldsymbol{p}_{i}}'-{\boldsymbol{p}_{i}} \right\|} \right)={\boldsymbol{M}_{{{h}_{i}}}}{\boldsymbol{c}_{i}}.
\end{eqnarray}
Let ${\boldsymbol{q}_{0}}\in R^2$ denotes the position of the target. 
If agent $i$ observe the target, 
agent $i$ can obtain the unit
bearing vector of target, which is given by
\begin{eqnarray}
    \label{model6}
    {{\boldsymbol{z}}_{i}}=\frac{{{\boldsymbol{r}}_{i}}}{\left\| {{\boldsymbol{r}}_{i}} \right\|}=\frac{{{\boldsymbol{q}}_{0}}-{{\boldsymbol{p}}_{i}}}{\left\| {{\boldsymbol{q}}_{0}}-{{\boldsymbol{p}}_{i}} \right\|}.
\end{eqnarray}
${{\boldsymbol{z}}_{i}}$ can also be expressed as
${{\boldsymbol{z}}_{i}}=[\cos\theta_i, \sin\theta_i ]^T$, 
where $\theta_i$ is the angle between ${\boldsymbol{r}}_{i}$ and $x$ axis. 
We consider that only two agents in the network can obtain the bearing information. Based on the differences in sensing, we divide agents into two types:
1) {sensing agent (SA)} $ (i\in {{\mathcal{I}}_{1}}\triangleq \left\{ 1,2 \right\})$: SAs can measure target bearing vector  ${{\boldsymbol{z}}_{i}}$;
2) {non-sensing agent (NSA)} $ (i\in {{\mathcal{I}}_{2}}\triangleq \left\{ 3,4,...,n \right\})$ : NSAs   can only exchange information with neighbors $N_i$.

Let ${\mathcal{\bar{G}}}$ denote the communication topology graph associated with $n + 1$ agents, including $n$ agents and a target. Note that ${\mathcal{\bar{G}}}$  consists of the following three layers:
\begin{itemize}
    \item \textbf{Target layer}: A target stays in this layer, and the connection between target and next layer is directed. 
    \item \textbf{SA layer}: Two SAs stay in this layer. The edge between these two SAs is undirected. 
    \item \textbf{NSA layer}: $n-2$ NSAs stay in this layer.  There is at least a directed path from SA layer to NSA layer. 
\end{itemize}

\begin{figure}
	\centering  
    \includegraphics[width=0.75\linewidth]{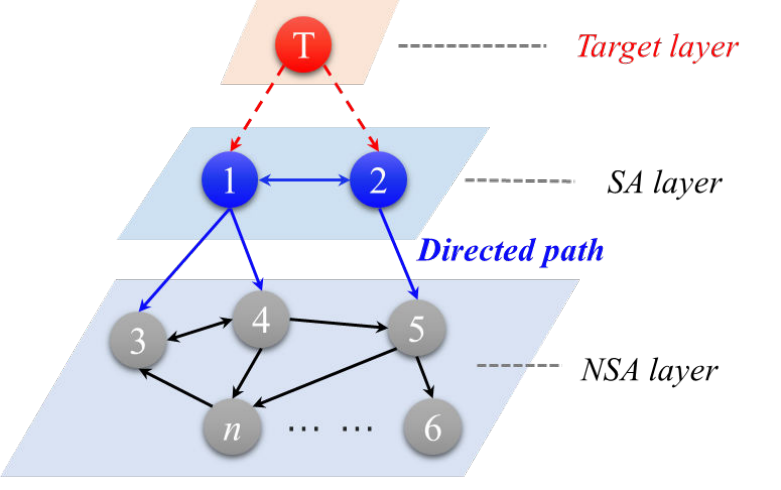}
	\caption{The connection between target, SAs and NSAs.}
    \label{pic_topol}
\end{figure}

An example of ${\mathcal{\bar{G}}}$ is displayed in Fig. \ref{pic_topol}.  Let  $\mathcal{G}^*$ denotes the topology graph of $n-2$ agents in NSA layer. $\boldsymbol{A}_{n} = [\alpha_{ij}]\in R^{(n-2)\times (n-2)}$ denotes the weighted adjacency matrix of the graph $\mathcal{G}^*$. 
$\alpha_{ii} = 0$ for all agents $(i\in \mathcal{I}_2)$, and $\alpha_{ij} > 0$ if agent $i$ can obtain information from its neighbor $j\in N_i$ and otherwise $\alpha_{ij} = 0$. 
The Laplacian matrix is defined as $\boldsymbol{L}\in R^{(n-2)\times (n-2)}$, where $l_{ii}= {\textstyle \sum_{j=3}^{n}} \alpha _{ij}$ and $l_{ij}=-\alpha _{ij}$, $j\in \mathcal{I}_2$. 
We employ input matrix $\boldsymbol{B}=diag(\beta_{3j},\beta_{4j},\ldots,\beta_{nj}), j\in \mathcal{I}_1$ to represent the connection between NSAs and SAs. 
In this paper, we provide the following assumption and the problem:

\begin{assumption}
\label{assumption_location}
SAs  are not collinear with the target.  
\end{assumption}

\begin{figure}
	\centering  
\subfigure[Initial state]{
    \includegraphics[width=0.4\linewidth]{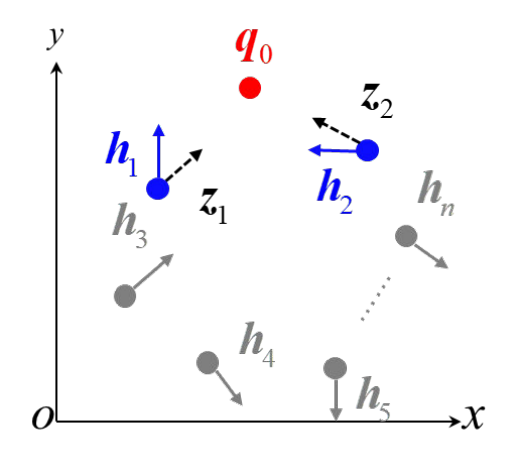}}
\subfigure[Ultimate state]{
    \includegraphics[width=0.4\linewidth]{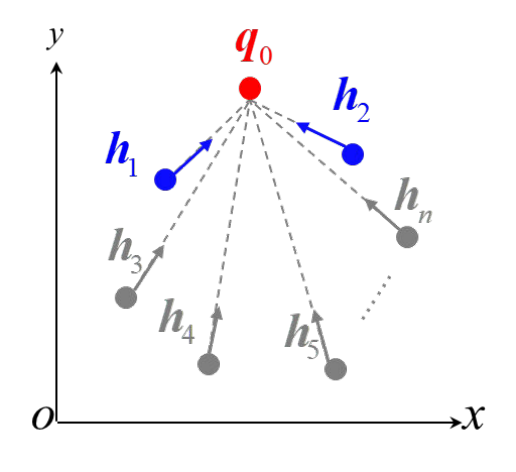}}
	\caption{The scenarios of target-pointing consensus.}
    \label{pic_problem}
\end{figure}

\begin{problem}
Given a $n$-agent system satisfying \textit{assumption 1},  design a distributed control algorithm to drive all the agents' headings to point at a target ${\boldsymbol{q}_{0}}$. The scenarios of target-pointing consensus are shown in Fig. \ref{pic_problem}.\end{problem}

\section{MAIN RESULTS}\label{section_m}

\subsection{ Bearing-only Estimator and Pointing Controller}

Our strategy consists of two parts: 1) a bearing-based estimator for each agent $i$ to estimate the target location ${\boldsymbol{q}_{0}}$, and 2) a controller for each agent to point at target.

Let each agent $i$ generate an estimate $\boldsymbol{\hat{q}}_i(t)$ of target location, and the initial estimate is $\boldsymbol{\hat{q}}_i(0)$. Based on the neighbor estimates $\boldsymbol{\hat{q}}_j$ and bearing vector $\boldsymbol{{h}}_j$, agent $i$ updates the estimate according to the following equations
\begin{subequations}
\begin{eqnarray}
    \label{model7}
{\boldsymbol{\dot{\hat{q}}}_{i}}={{{k }_{ij}}\left( {{\boldsymbol{\hat{q}}}_{j}}-{{\boldsymbol{\hat{q}}}_{i}} \right)-{\boldsymbol{M}_{{{z}_{i}}}}\left( {{\boldsymbol{\hat{q}}}_{i}}-{\boldsymbol{p}_{i}} \right),i\in {{\mathcal{I}}_{1}},}
\end{eqnarray}

    \begin{eqnarray}
    \label{model8}
{\boldsymbol{\dot{\hat{q}}}_{i}}=
   \sum\limits_{j\in {{N }_{i}^{n}}}{{{\alpha }_{ij}}\left( {{\boldsymbol{\hat{q}}}_{j}}-{{\boldsymbol{\hat{q}}}_{i}}\right)+
   \sum\limits_{j\in {{N }_{i}^{s}}}{{\beta }_{ij}}\left( {{\boldsymbol{\hat{q}}}_{j}}-{{\boldsymbol{\hat{q}}}_{i}}\right),i\in {{\mathcal{I}}_{2}},  }
\end{eqnarray}
\end{subequations}
where  Eq. (\ref{model7}) is the estimator for SAs, while Eq. (\ref{model8}) is provided for NSAs. $k_{ij}$ denotes the parameter of the SA estimator, and   
$\boldsymbol{M}_{{{z}_{i}}}$ represents an orthogonal projection matrix corresponding to the bearing vector $\boldsymbol{z}_{i}$, which is given by
\begin{eqnarray}\label{model9}
\boldsymbol{M}_{{{z}_{i}}}={\boldsymbol{I}_{2}}-{\boldsymbol{z}_{i}}{\boldsymbol{z}_{i}}^{T}.
\end{eqnarray}
The bearing-based estimator (\ref{model8}) consists of two terms:

\begin{itemize}

\item \textbf{Consensus term}: $-\sum{{{\alpha }_{ij}}\left( {{\boldsymbol{\hat{q}}}_{j}}-{{\boldsymbol{\hat{q}}}_{i}} \right)}$ is the consensus term that guides the network to maintain estimation alignment during the transients. This term affects the change of  $\boldsymbol{\hat{q}}_i(t)$ as long as ${{\boldsymbol{\hat{q}}}_{j}} \ne {{\boldsymbol{\hat{q}}}_{i}} $.

\item \textbf{Orthogonal projection term}: ${{\beta }_{i}}{\boldsymbol{M}_{{{z}_{i}}}}\left( {{\boldsymbol{\hat{q}}}_{i}}-{\boldsymbol{p}_{i}} \right)$ is an adjustment term that make $\boldsymbol{\hat{q}}_i(t)$ move perpendicularly to the line passing through the agent and the target. Note that
if vectors $\boldsymbol{z}_{i}$ and $(\boldsymbol{\hat{q}}_i-\boldsymbol{p}_i)$ are parallel, or $\boldsymbol{z}_{i}\times(\boldsymbol{\hat{q}}_i-\boldsymbol{p}_i)=\boldsymbol{0}$, this term vanishes. So the effect of this item is to always maintain the $\boldsymbol{\hat{q}}_i$ on the line between $\boldsymbol{{q}}_0$ and $\boldsymbol{{p}}_i$.
\end{itemize}
In addition, we can find that the neighbors of the two items in Eq. (\ref{model8})  are different. The first term of Eq. (\ref{model8}) denotes the interaction of NSAs, and the second term denotes the impact of SAs on NSAs.

After target localization, agents need to drive the headings $\boldsymbol{h}_i$ to point at the target based on $\boldsymbol{\hat{q}}_i$. For each agent $i \in \mathcal{V}$, the pointing controller is designed as follow:
\begin{eqnarray}\label{model10}
\boldsymbol{{\dot{h}}}_i = \boldsymbol{{M}}_{h_i}\left(\boldsymbol{\hat{q}}_i-\boldsymbol{p}_i\right),
\end{eqnarray}
where the matrix $\boldsymbol{{M}}_{h_i}$ have been given in Eq. (\ref{model3}).

\subsection{Localizability and Stability Analysis}

In this section, we will analyze 1) the localizablility of location estimators, and 2) the stability of the closed-loop system. 
We derive the following equations based on (\ref{model7}), 
\begin{subequations}

\begin{eqnarray}
    \label{model12}
{\boldsymbol{\dot{\hat{q}}}_{1}}={k_{12}}\left( {{\boldsymbol{\hat{q}}}_{2}}-{{\boldsymbol{\hat{q}}}_{1}} \right)-{\boldsymbol{M}_{{{z}_{1}}}}\left( {{\boldsymbol{\hat{q}}}_{1}}-{\boldsymbol{p}_{1}} \right),
\end{eqnarray}

\begin{eqnarray}
    \label{model13}
{\boldsymbol{\dot{\hat{q}}}_{2}}={{k }_{21}}\left( {{\boldsymbol{\hat{q}}}_{1}}-{{\boldsymbol{\hat{q}}}_{2}} \right)-{\boldsymbol{M}_{{{z}_{2}}}}\left( {{\boldsymbol{\hat{q}}}_{2}}-{\boldsymbol{p}_{2}} \right),
\end{eqnarray}
\end{subequations}
where $k_{12}>0$, ${k }_{21}>0$. Since ${{\boldsymbol{z}}_{i}}=[\cos\theta_i, \sin\theta_i ]^T$, we have
\begin{eqnarray}
    \label{model14}
{\boldsymbol{M}_{{{z}_{i}}}}=\left[ \begin{matrix}
   {{\sin }^{2}}{{\theta }_{i}} & -\cos {{\theta }_{i}}\sin {{\theta }_{i}}  \\
   -\cos {{\theta }_{i}}\sin {{\theta }_{i}} & {{\cos }^{2}}{{\theta }_{i}}  \\
\end{matrix} \right].
\end{eqnarray}
The estimation error ${\boldsymbol{\tilde{q}}_{i}}={\boldsymbol{q}_{0}}-\boldsymbol{\hat{q}}_i$ is introduced, and the derivative of ${\boldsymbol{\tilde{q}}_{i}}$ is given by
\begin{eqnarray}
    \label{model15}
{\boldsymbol{\dot{\tilde{q}}}_{i}}=-\boldsymbol{\dot{\hat{q}}}_i,
\end{eqnarray}

Substitute Eqs. (\ref{model12}) and (\ref{model13}) into Eq. (\ref{model15}), we have

\begin{subequations}\label{model1611}
\begin{eqnarray}
\label{model161}
   {{\boldsymbol{\dot{\tilde{q}}}}_{1}}=-{k_{12}}( {{\boldsymbol{\tilde{q}}}_{1}}-{{\boldsymbol{\tilde{q}}}_{2}} )-{\boldsymbol{M}_{{{z}_{1}}}}{{\boldsymbol{\tilde{q}}}_{1}}, 
   \end{eqnarray}
   \begin{eqnarray}
\label{model171}
  {{\boldsymbol{\dot{\tilde{q}}}}_{2}}=-{{k }_{21}}( {{\boldsymbol{\tilde{q}}}_{2}}-{{\boldsymbol{\tilde{q}}}_{1}} )-{\boldsymbol{M}_{{{z}_{2}}}}{{\boldsymbol{\tilde{q}}}_{2}}.
  \end{eqnarray}
\end{subequations}

\bigskip
\begin{lemma}\label{lemma_SA}
Under \emph{Assumption \ref{assumption_location}}, estimation errors ${{\boldsymbol{{\tilde{q}}}}_{1}}$ and ${{\boldsymbol{{\tilde{q}}}}_{2}}$ exponentially converge to $\boldsymbol{0}$ as $t\to \infty$. 
\end{lemma}

\emph{Proof}: 
The vector form of Eq. (\ref{model1611}) is given by
\begin{eqnarray}
    \label{model18}
    \boldsymbol{\dot{\tilde{q}}}\left( t \right)=\boldsymbol{H}\boldsymbol{\tilde{q}}\left( t \right),
\end{eqnarray}
where $\boldsymbol{\dot{\tilde{q}}}\left( t \right)=\left[\boldsymbol{\dot{\tilde{q}}_1}^T, \boldsymbol{\dot{\tilde{q}}_2}^T\right]^T$. 
Note that ${\theta }_{1}$ and ${\theta }_{2}$ are constants since the target is stationary. Thus the system (\ref{model18}) is a linear time-invariant (LTI) system. 
The characteristic polynomial of ${\boldsymbol{H}}$ is 
\begin{equation}\label{model23}
 \begin{aligned}
  f(\lambda)=& {{\lambda }^{4}}+(2{k_{12}}+2{{k }_{21}}+2){{\lambda }^{3}} \\ 
 & +({k_{12}}^{2}+{{k }_{21}}^{2}+2{k_{12}}{{k }_{21}}+3{k_{12}}+3{{k }_{21}}+1){{\lambda }^{2}} \\ 
 & +({{k }_{21}}^{2}+{k_{12}}^{2}+2{k_{12}}{{k }_{21}}+{{k }_{21}}+{k_{12}}) \lambda\\ 
 & +{k_{12}}{{k }_{21}}{{\sin }^{2}}({{\theta }_{1}}-{{\theta }_{2}}), \\ 
\end{aligned}
\end{equation}
 where coefficients of $f(\lambda)$ are all positive real numbers. According to Hurwitz criterion, it is easy to verify that the necessary and sufficient condition for system stability is $\sin \left( {{\theta}_{1}}-{{\theta}_{2}} \right)\ne 0$. The stability of the system is equivalent to the exponential convergence of ${\boldsymbol{\tilde{q}}}_{1}$ and ${\boldsymbol{\tilde{q}}}_{2}$ in LTI system.  We obtain the condition for asymptotic convergence by solving $\sin \left( {{\theta}_{1}}-{{\theta}_{2}} \right)\ne 0$,
 \begin{eqnarray} 
     \label{model24}
     \left\{ \begin{matrix}
  {{\theta}_{2}}\ne {{\theta}_{1}}, \\ 
  {{\theta}_{2}}\ne {{\theta}_{1}}+\pi . \\ 
\end{matrix} \right.
 \end{eqnarray}
 Condition  (\ref{model24}) is equivalent to \emph{Assmption 1}. In another word, the bearing-only localizablility of the target is naturally guaranteed if \emph{Assmptions 1} is met. So ${\boldsymbol{\tilde{q}}}_{1}$ and ${\boldsymbol{\tilde{q}}}_{2}$  converge exponentially to the origin. This ends the proof. 

 \bigskip
\begin{figure}[H]
	\centering  
    \includegraphics[width=0.8\linewidth]{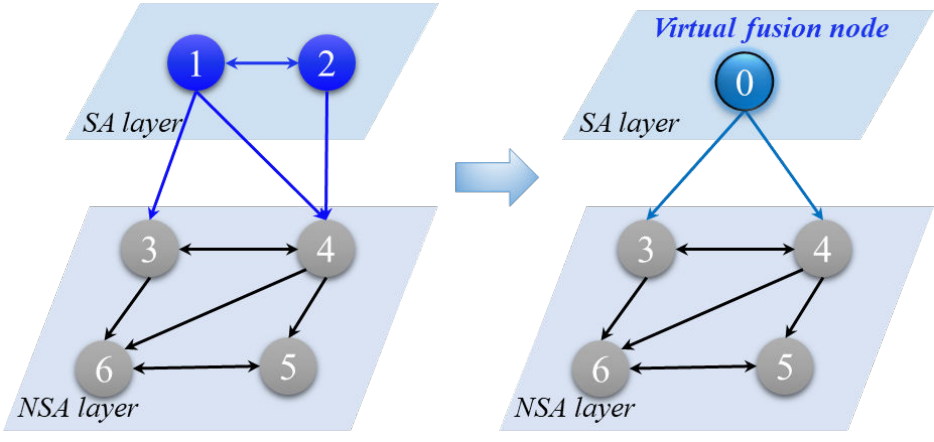}
	\caption{The graph using virtual fusion node 0.}
    \label{pic_fusion}
\end{figure}

Next, we give the localization analysis of the NSAs.
In this section, the concept of virtual fusion node is introduced to analyze the convergence of NSAs' estimates. The actual connectivity is transformed into the a new topology graph consisting of $n-2$ NSAs and a virtual fusion node.

In Fig. \ref{pic_topol}, SAs can be regarded as leaders of NSAs,  and the estimates of SAs converge to a constant vector according to \emph{Lemma \ref{lemma_SA}}. So the problem 2) can be converted into a consensus tracking problem of NSA layer. The estimator (\ref{model8}) can be written as
 \begin{equation} 
     \label{model25}
  \begin{aligned}
  {{{\boldsymbol{\dot{\hat{q}}}}}_{i}}= &\sum\limits_{j\in N_{i}^{n}}{{{\alpha }_{ij}}\left( {{{\boldsymbol{\hat{q}}}}_{j}}-{{{\boldsymbol{\hat{q}}}}_{i}} \right)} \\ 
 & +{{\beta }_{i1}}\left( {{{\boldsymbol{\hat{q}}}}_{1}}-{{{\boldsymbol{\hat{q}}}}_{i}} \right)+{{\beta }_{i2}}\left( {{{\boldsymbol{\hat{q}}}}_{2}}-{{{\boldsymbol{\hat{q}}}}_{i}} \right). 
\end{aligned}
 \end{equation}
Substituting errors  ${{{\boldsymbol{\tilde{q}}}}_{i}}$ and ${{\boldsymbol{\dot{\tilde{q}}}}_{i}}$ into Eq. (\ref{model25}) yields
 \begin{equation}
   \label{model26}
     \begin{aligned}
  {{{\boldsymbol{\dot{\tilde{q}}}}}_{i}}=& -\sum\limits_{j\in N_{i}^{n}}{{{\alpha }_{ij}}\left( {{{\boldsymbol{\tilde{q}}}}_{i}}-{{{\boldsymbol{\tilde{q}}}}_{j}} \right)} \\ 
 & -{{\beta}_{i1}}\left( {{{\boldsymbol{\tilde{q}}}}_{i}}-{{{\boldsymbol{\tilde{q}}}}_{1}} \right)-{{\beta }_{i2}}\left( {{{\boldsymbol{\tilde{q}}}}_{i}}-{{{\boldsymbol{\tilde{q}}}}_{2}} \right) . 
\end{aligned}
 \end{equation} 
By merging the last two items, Eq. (\ref{model26}) can be written as
 \begin{equation}
     \label{model261}
     \begin{aligned}
{{\boldsymbol{\dot{\tilde{q}}}}_{i}}=&-\sum\limits_{j\in N_{i}^{n}}{{{\alpha }_{ij}}\left( {{{\boldsymbol{\tilde{q}}}}_{i}}-{{{\boldsymbol{\tilde{q}}}}_{j}} \right)}\\
&-\left( {{\beta }_{i1}}+{{\beta }_{i2}} \right)\left( {{{\boldsymbol{\tilde{q}}}}_{i}}-{{{\boldsymbol{\tilde{q}}}}_{1}} \right)-{{\beta }_{i2}}\left( {{{\boldsymbol{\tilde{q}}}}_{1}}-{{{\boldsymbol{\tilde{q}}}}_{2}} \right).
\end{aligned}
 \end{equation}
 By the above operation, we successfully merge agent 1 and agent 2 into a new agent with a coefficient $\left( {{\beta }_{i1}}+{{\beta }_{i2}} \right)$. In the network, We regard this agent as the virtual fusion node 0. The new graph $\tilde{\mathcal{G}}$  composed of NSAs and virtual fusion node is a directed graph, which is illustrated in Fig. \ref{pic_fusion}. 
 To analyze the convergence of model (\ref{model261}), we provide the following assumption:

\begin{assumption}
\label{assumption_topology}
$\exists \; i\in \mathcal{I}_2$, ${{\beta }_{i1}}+{{\beta }_{i2}} >0$. 
And the directed topology graph $\tilde{\mathcal{G}}$ contains a spanning tree. Each node of NSA layer can find a directed path from fusion node 0 to itself. 
\end{assumption}

Based on \emph{Assumption \ref{assumption_topology}}, we can conclude the following theorem:

\begin{theorem}
    \label{theorem_location}
    Under \emph{Assumptions \ref{assumption_location} -- \ref{assumption_topology}},  $\forall  i \in \mathcal{I}$, $\boldsymbol{\tilde{q}}_i$
 converge exponentially to the origin as $t\to \infty$.
\end{theorem}

\emph{Proof}:
 The vector form of Eq. (\ref{model261}) is:
 \begin{equation}
     \label{model27}
     \begin{aligned}
     \boldsymbol{\dot{\tilde{q}}}^*=&-\left[(\boldsymbol{L}+\boldsymbol{\bar{B}}_f)\otimes \boldsymbol{I}_2 \right]\boldsymbol{{{\tilde{q}}}}^*\\
     &+\underbrace{(\boldsymbol{B}_f\otimes \boldsymbol{I}_2 )\boldsymbol{\tilde{q}}_1}_{\boldsymbol{\gamma }}-\underbrace{(\boldsymbol{B}_e\otimes \boldsymbol{I}_2 )(\boldsymbol{\tilde{q}}_1-\boldsymbol{\tilde{q}}_2)}_{\boldsymbol{\delta}  },
  \end{aligned}
 \end{equation}
 where $\boldsymbol{{\tilde{q}}}^*=\left[\boldsymbol{{\tilde{q}}}_3^T, \boldsymbol{{\tilde{q}}}_4^T,...,\boldsymbol{{\tilde{q}}}_n^T\right]^T$,  $\boldsymbol{B}_f=[\beta_{31}+\beta_{32}, \beta_{41}+\beta_{42}, ..., \beta_{n1}+\beta_{n2}]^T$,  $\boldsymbol{B}_e=[\beta_{32}, \beta_{42}, ..., \beta_{n2}]^T$ and $\boldsymbol{\bar{B}}_f=diag(\boldsymbol{{B}}_f)$. $\boldsymbol{I}_2 \in R^{2\times2}$ denotes a unit matrix, and $\otimes$ is the Kronecker product. 
$(\boldsymbol{\gamma}-\boldsymbol{\delta})$ can be regarded as an external signal to model (\ref{model27}). 

 According to \emph{ Theorem 3.6} in \cite{ren2005consensus}, $\left[(\boldsymbol{L}+\boldsymbol{\bar{B}}_f)\otimes \boldsymbol{I}_2 \right]$ is a positive-definite matrix under \emph{Assumption \ref{assumption_topology}}. So the model (\ref{model27}) is input-to-state stable (ISS). According to \emph{Lemma \ref{lemma_SA}}, signals $\boldsymbol{{\tilde{q}}}_1$ and $\boldsymbol{{\tilde{q}}}_2$ will  converge exponentially to $\boldsymbol{0}$. As a result, $(\boldsymbol{\gamma}-\boldsymbol{\delta}) \to \boldsymbol{0}$ exponentially fast. 
 The exponential convergence of $\boldsymbol{{\tilde{q}}}^*$ can also be concluded. This ends the proof.

\bigskip
\begin{theorem}
    \label{theorem_control}
    Suppose that \emph{Assumptions \ref{assumption_location} -- \ref{assumption_topology}} hold. Under strategies (\ref{model7}) -- (\ref{model10}), all agents’ headings asymptotically point toward the target $\boldsymbol{q}_0$. 
\end{theorem}

 \emph{Proof}: See \emph{Theorem 3.4} in \cite{Trinh2020}.

\begin{figure}[H]
	\centering  
\subfigure[Estimation errors]{
    \includegraphics[width=0.45\linewidth]{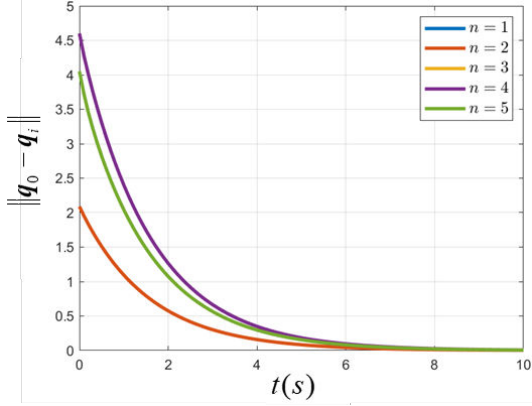}}
\subfigure[Pointing errors]{
    \includegraphics[width=0.45\linewidth]{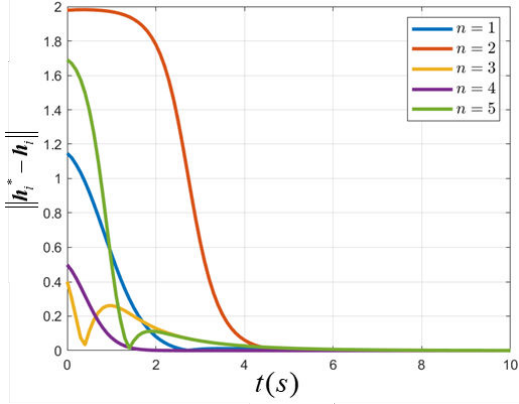}}
	\caption{The trajectories of estimation and pointing errors.}
    \label{pic_error}
\end{figure}

\section{Simulations and CONCLUSIONS}\label{section_s}

\begin{figure}[H]
	\centering  
\subfigure[Initial states]{
    \includegraphics[width=0.475\linewidth]{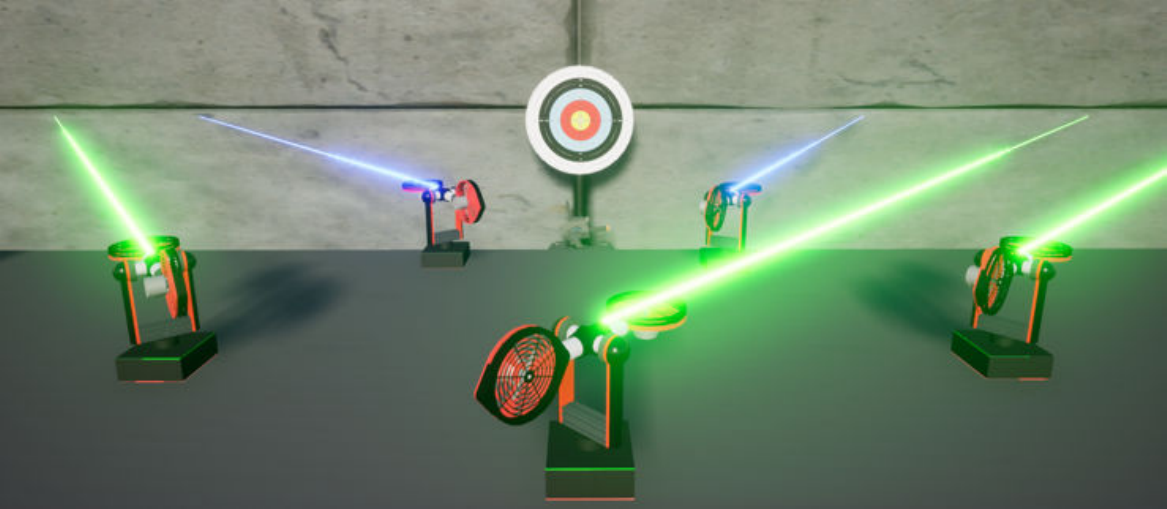}}
\subfigure[Ultimate states]{
    \includegraphics[width=0.46\linewidth]{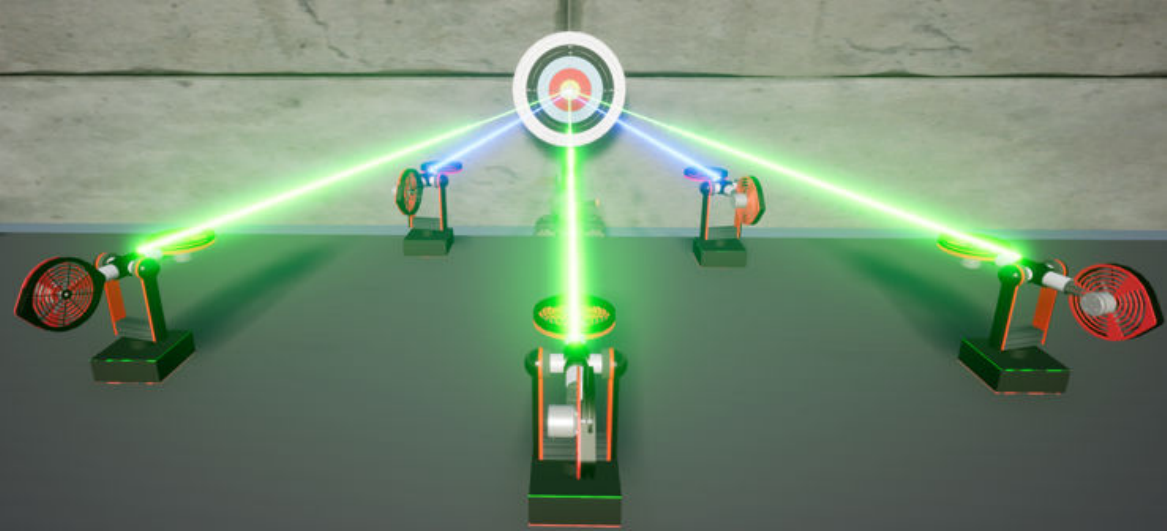}}
	\caption{The convergence procedure of 6 agents in UE4.}
    \label{pic_ue4}
\end{figure}

Supposing \emph{Assumption \ref{assumption_topology}} are satisfied, a six-agent system with 2 SAs and 4 NSAs is considered. 
Six agents are positioned at $[2,4]^T$, $[4,4]^T$, $[5,2]^T$, $[4,1]^T$, $[2,1]^T$ and $[1,2]^T$. The target is positioned at $[3,6]^T$.
The connection of these agents is illustrated in Fig. \ref{pic_fusion}. 
The initial heading vectors $\boldsymbol{h}_i(0)$ are randomly chosen. The initial esitimates $\boldsymbol{\hat{q}}_i(0)$ are chosen as same as their positions $\boldsymbol{p}_i$. We simulate the six-agent system under strategies (\ref{model7}) - (\ref{model10}) with the simulation step $\tau=0.1s$. 
The error trajectories and UE4 simulation results of pointing platforms are shown in Fig. \ref{pic_error} and Fig. Fig. \ref{pic_ue4}. It is seen that all agents point at the target eventually. Fig. \ref{pic_error} shows that both estimation errors and pointing errors converge asymptotically to zeros. The effectiveness of our strategies is verified by this simulation case.

In this paper, in order to achieve target-pointing consensus, we have constructed a two-step strategy containing a bearing-only estimator
for target location and a control law for  target pointing.  The localizability and stability of the closed-loop system have been analyzed based on a hierarchical graph with three layers.

\end{document}